\documentclass[11pt,a4paper]{article}
\usepackage[dvips]{graphics}
\usepackage{epsfig}
\usepackage{amssymb,amsfonts,amsmath}

\begin{document}

\title{A stochastic approach to the solution of magnetohydrodynamic equations%
}
\author{E. Floriani\thanks{%
Centre de Physique Th\'{e}orique, CNRS, Luminy - Marseille,
floriani@cpt.univ-mrs.fr} \ \ and R. Vilela Mendes\thanks{%
CMAF - Complexo Interdisciplinar, Universidade de Lisboa Av. Gama Pinto 2,
1649-003 Lisboa, Portugal, vilela@cii.fc.ul.pt, } \thanks{%
Instituto de Plasmas e Fus\~{a}o Nuclear, IST, vilela.mendes@ist.utl.pt}}
\date{ }
\maketitle

\begin{abstract}
The construction of stochastic solutions is a powerful method to obtain
localized solutions in configuration or Fourier space and for parallel
computation with domain decomposition. Here a stochastic solution is
obtained for the magnetohydrodynamics equations. Some details are given
concerning the numerical implementation of the solution which is illustrated
by an example of generation of long-range magnetic fields by a velocity
source.
\end{abstract}

Keywords: Kinetic equations, Stochastic solutions, Magnetohydrodynamics

\section{Introduction}

To solve complex differential problems in large domains, one way to profit
from parallel computation in multiprocessor machines is to decompose the
domain into subdomains and assign the problem in each subdomain to a
different processor. However, one is left with the problem of computation of
the boundary conditions in the subdomain interfaces. This implies that, in
addition to the time consumed solving the equation in each subdomain, a
considerable amount of time will also be consumed in the communication
between processors. The ideal situation would be to have a method to compute
local solutions at the interface points without the need for a grid. Such a
method is implicit in the idea of stochastic solutions.

In the following the notion of stochastic solution is introduced for linear
partial differential equations and a method is described that extends this
notion to nonlinear equations. In addition to its use in domain
decomposition problems, other relevant features of the stochastic solutions
are discussed.

\subsection{The notion of stochastic solution: Linear and nonlinear partial
differential equations}

Linear elliptic and parabolic equations (both with Cauchy and Dirichlet
boundary conditions) have a probabilistic interpretation. This is a
classical result and a standard tool in potential theory. As a simple
example consider the heat equation%
\begin{equation}
\frac{\partial}{\partial t}u(t,x)=\frac{1}{2}\frac{\partial ^{2}}{\partial
x^{2}}u(t,x)\qquad \text{with}\qquad u(0,x)=f(x)  \label{1.1a}
\end{equation}%
The solution may be written either as 
\begin{equation}
u\left( t,x\right) =\frac{1}{\sqrt{2\pi }}\int \frac{1}{\sqrt{t}}\exp \left(
-\frac{\left( x-y\right) ^{2}}{2t}\right) f\left( y\right) dy  \label{1.2}
\end{equation}%
or as 
\begin{equation}
u(t,x)=\mathbb{E}_{x}f(X_{t})  \label{1.3a}
\end{equation}%
$\mathbb{E}_{x}$ being the expectation value, starting from $x$, of the
Wiener process $dX_{t}=dW_{t}$.

Whereas Eq.(\ref{1.1a}) is a \textit{specification} of the problem, Eqs.(\ref%
{1.2}) and (\ref{1.3a}) are \textit{solutions} in the sense that they both
provide algorithms for the construction of a function satisfying the
specification. An important condition for (\ref{1.2}) and (\ref{1.3a}) to be
considered as \textit{solutions} is the fact that the algorithm is \textit{%
independent} of the particular solution, in the first case an \textit{%
integration} procedure and in the second a \textit{solution-independent}%
\textbf{\ }process. This should be contrasted with stochastic processes
constructed from a given particular solution, as has been done, for example,
for the Boltzmann equation.

Whenever a similar stochastic process algorithm may be associated to
nonlinear equations, this would provide new \textit{exact solutions} and new 
\textit{numerical algorithms. }To obtain stochastic solutions for nonlinear
equations, it is useful to recall that in the linear partial differential
equation case the stochastic process starts from the point $x$ where the
solution is to be computed and the solution is a functional of the exit
values of the process (from a space domain $D$ or a space-time domain $%
D\times \left[ 0,t\right] $). Therefore, it is natural to conjecture that
for the nonlinear equations the relevant process would have a diffusion,
propagation or jump component associated to the linear part of the equation
plus a branching mechanism associated to the nonlinear part. Then the
solution would also be a functional of the exit measures generated by the
process.

For the implementation of this conjecture one rewrites the equation as an
integral one, for which a probabilistic interpretation is given. In the end
the stochastic solution is the expectation of a functional over a
tree-indexed measure. The method, which leads to rigorous results, may also
be looked at, in qualitative terms, as importance sampling evaluation of the
Picard series.

This method was first used in the pioneering paper of McKean \cite{McKean}
for the Kolmogorov-Petrovskii-Piskonov (KPP)\ equation. Later, a similar
technique was used for the Navier-Stokes \cite{LeJan} \cite{Waymire}, the
Vlasov-Poisson \cite{Cipriano} \cite{Floriani} \cite{Vilela2} \cite{Vilela3}%
, the Euler \cite{Vilela1} and a fractional version of the KPP equation \cite%
{Ouerdiane}. For the diffusion equation with $u^{\alpha }$ $\left( \alpha
\in \left[ 0,2\right] \right) $ nonlinearities, Dynkin uses a different
method \cite{Dynkin}, namely scaling limits leading to superprocesses (for a
comparison of the McKean-type construction and superprocesses see \cite%
{Vilela4}).

\subsection{Stochastic solutions and numerical algorithms}

Once a stochastic solution is obtained for a partial differential equation,
how does it stand in comparison with deterministic numerical algorithms? The
main points to be considered are:

(a) Stochastic solutions may provide new exact solutions in cases where
exact solutions were not known before.

(b) Deterministic algorithms grow exponentially with the dimension $d$ of
the space, roughly $N^{d}$ ($\frac{L}{N}$ being the linear size of the grid)
whereas the numerical implementation of a stochastic process only grows with
the dimension $d$.

(c) Deterministic algorithms aim at obtaining the solution in the whole
domain. Then, even if an efficient deterministic algorithm exists, the
stochastic algorithm might still be competitive if only localized values of
the solution are desired. For example by studying only a few high Fourier
modes one may obtain information on the small scale fluctuations which would
require a very fine grid in a deterministic algorithm.

(d) Each sample path of the stochastic process is independent of the others.
Likewise, paths starting from different points are independent from each
other. Therefore the stochastic algorithms are a natural choice for parallel
and distributed computation.

(e) Stochastic algorithms handle equally well regular and complex boundary
conditions although, of course, the computation of exit times from complex
domains might not be an easy matter.

(f) Also, as already pointed out, the local nature of the stochastic
solutions make them the most appropriate choice to obtain boundary
conditions in subdomain interfaces, thus avoiding the time consuming
communication problem. The remarkable efficiency of this method for domain
decomposition schemes has been described in \cite{Acebron1} \cite{Acebron2} 
\cite{Acebron3}.

\section{Stochastic solutions of magnetohydrodynamics equations}

Magnetohydrodynamics concerns the dynamics of magnetic fields in
electrically conducting fluids, e.g. plasmas or liquid metals. It is a
macroscopic theory which may be considered as an approximation of the
Boltzmann's equation when the space and time scales are larger than all
relevant length scales, such as the Debye length or the gyro-radius of the
charged particles. We consider, in a non-relativistic approximation, the
equations of magnetohydrodynamics in 3 dimensions, with non-zero fluid
viscosity and electric resistivity. The fluid is taken to be incompressible
with density $\rho (x,t)=\rho _{0}$ constant and uniform. The equations for
the velocity $\overset{\rightarrow }{V}(x,t)$ of the fluid and the magnetic
field $\overset{\rightarrow }{B}(x,t)$ are:%
\begin{equation}
\frac{\partial \overset{\rightarrow }{V}}{\partial t}=-(\overset{\rightarrow 
}{V}\cdot \nabla )\overset{\rightarrow }{V}+\frac{1}{\rho _{0}\mu _{0}}(%
\overset{\rightarrow }{B}\cdot \nabla )\overset{\rightarrow }{B}-\frac{1}{%
2\rho _{0}\mu _{0}}\nabla \overset{\rightarrow }{B}^{2}-\frac{1}{\rho _{0}}%
\nabla P+\nu \nabla ^{2}\overset{\rightarrow }{V}+\overset{\rightarrow }{F}%
(x,t)  \label{dvdt}
\end{equation}%
\begin{equation}
\frac{\partial \overset{\rightarrow }{B}}{\partial t}=-(\overset{\rightarrow 
}{V}\cdot \nabla )\overset{\rightarrow }{B}+(\overset{\rightarrow }{B}\cdot
\nabla )\overset{\rightarrow }{V}+\frac{\eta }{\mu _{0}}\nabla ^{2}\overset{%
\rightarrow }{B}  \label{dbdt}
\end{equation}%
$\nu $ being the kinematic viscosity, $\eta $ the resistivity and $\mu _{0}$
the vacuum permeability. $\overset{\rightarrow }{F}(x,t)$ is a forcing term
for the fluid velocity. For the Fourier transformed quantities,%
\begin{eqnarray*}
\overset{\rightarrow }{v}(k,t) &=&(2\pi )^{-3/2}\int d^{3}x\,\overset{%
\rightarrow }{V}(x,t)\,e^{ik\cdot x} \\
\overset{\rightarrow }{b}(k,t) &=&(2\pi )^{-3/2}\int d^{3}x\,\overset{%
\rightarrow }{B}(x,t)\,e^{ik\cdot x} \\
\overset{\rightarrow }{f}(k,t) &=&(2\pi )^{-3/2}\int d^{3}x\,\overset{%
\rightarrow }{F}(x,t)\,e^{ik\cdot x}
\end{eqnarray*}%
use the fact that the divergences of $\overset{\rightarrow }{V}(x,t),\overset%
{\rightarrow }{B}(x,t),\overset{\rightarrow }{F}(x,t)$ vanish and project on
the plane orthogonal to the vector $k$. The projection operator is 
\begin{equation*}
\pi _{(k)}(\overset{\rightarrow }{\xi })=\overset{\rightarrow }{\xi }-(%
\overset{\rightarrow }{\xi }\cdot \overset{\rightarrow }{e_{k}})\overset{%
\rightarrow }{e_{k}}
\end{equation*}%
with $e_{k}=\frac{k}{|k|}$. The projection eliminates the gradient terms
and, since $k\cdot v(k,t)=k\cdot b(k,t)=0$, no information is lost on the
velocity and magnetic fields. Then we obtain%
\begin{eqnarray}
\frac{\partial \overset{\rightarrow }{v}(k,t)}{\partial t} &=&-\nu k^{2}%
\overset{\rightarrow }{v}(k,t)+(2\pi )^{-3/2}|k|\,\int d^{3}q\,\left\{ 
\overset{\rightarrow }{v}(q,t)\otimes _{k}\overset{\rightarrow }{v}%
(k-q,t)\right.  \notag \\
&&\left. -\frac{1}{\rho _{0}\mu _{0}}\,\overset{\rightarrow }{b}(q,t)\otimes
_{k}\overset{\rightarrow }{b}(k-q,t)\right\} +\overset{\rightarrow }{\varphi 
}(k,t)  \label{piv}
\end{eqnarray}%
\begin{eqnarray}
\frac{\partial \overset{\rightarrow }{b}(k,t)}{\partial t} &=&-\,\frac{\eta 
}{\mu _{0}}k^{2}\overset{\rightarrow }{b}(k,t)+(2\pi )^{-3/2}|k|\,\int
d^{3}q\,\left\{ \overset{\rightarrow }{v}(q,t)\otimes _{k}\overset{%
\rightarrow }{b}(k-q,t)\right.  \notag \\
&&\left. -\overset{\rightarrow }{b}(q,t)\otimes _{k}\overset{\rightarrow }{v}%
(k-q,t)\right\}  \label{pib}
\end{eqnarray}%
where $\overset{\rightarrow }{\varphi }(k,t)=\pi _{(k)}\overset{\rightarrow }%
{f}(k,t)$ is the Fourier transform of the divergenceless part of the forcing
and the product $\otimes _{k}$ between two vectors $\overset{\rightarrow }{%
\xi },\overset{\rightarrow }{\omega }$ is defined by 
\begin{equation}
\overset{\rightarrow }{\xi }\otimes _{k}\overset{\rightarrow }{\omega }=i(%
\overset{\rightarrow }{e_{k}}\cdot \overset{\rightarrow }{\xi })\,\pi _{(k)}%
\overset{\rightarrow }{\omega }  \label{3.7}
\end{equation}

The next step will be to give a probabilistic interpretation to the
magnetohydrodynamics equations by defining a process and an associated
functional that provides the solution. Two constructions will be given. They
both lead to rigorous results. However, for practical purposes and numerical
implementation the one that is most convenient will depend on the values of
the equation physical parameters.

\subsection{Dissipation-controlled stochastic clock}

To guarantee convergence of the functionals associated to the stochastic
processes it is convenient to rescale the vectors $\overset{\rightarrow }{v}%
(k,t)$ and $\overset{\rightarrow }{b}(k,t)$ by a $k-$dependent function $%
h(k) $. This rescaling should be familiar from the convergence proofs of
Picard iteration and will also be called \textit{the majorizing kernel}.
Define:

\begin{equation}
\overset{\rightarrow }{\chi _{v}}(k,t)=\frac{\overset{\rightarrow }{v}(k,t)}{%
h(k)}\;,\;\;\;\;\;\;\overset{\rightarrow }{\chi _{b}}(k,t)=\frac{\overset{%
\rightarrow }{b}(k,t)}{\sqrt{\rho _{0}\mu _{0}}\;h(k)}  \label{3.8}
\end{equation}%
Then the integral equations equivalent to (\ref{piv}-\ref{pib}) are:%
\begin{eqnarray}
\overset{\rightarrow }{\chi _{v}}(k,t) &=&\overset{\rightarrow }{\chi _{v}}%
(k,0)e^{-\nu k^{2}t}+\int_{0}^{+\infty }ds\,\nu k^{2}\,e^{-\nu
k^{2}s}\left\{ \frac{1}{3}\overset{\rightarrow }{\rho }(k,t-s)+\int d^{3}q%
\frac{h(q)h(k-q)}{(h\ast h)(k)}\right.  \notag \\
&&\left[ \frac{1}{3}\,g_{v\rightarrow vv}(k)\,\overset{\rightarrow }{\chi
_{v}}(q,t-s)\otimes _{k}\overset{\rightarrow }{\chi _{v}}(k-q,t-s)+\right. 
\notag \\
&&\left. \left. \frac{1}{3}\,g_{v\rightarrow bb}(k)\overset{\rightarrow }{%
\chi _{b}}(q,t-s)\otimes _{k}\overset{\rightarrow }{\chi _{b}}(k-q,t-s)%
\right] \right\}  \label{chiv}
\end{eqnarray}%
and%
\begin{eqnarray}
\overset{\rightarrow }{\chi _{b}}(k,t) &=&\overset{\rightarrow }{\chi _{b}}%
(k,0)e^{-\frac{\eta }{\mu _{0}}\,k^{2}t}+\int_{0}^{+\infty }ds\,\frac{\eta }{%
\mu _{0}}\,k^{2}\,e^{-\frac{\eta }{\mu _{0}}\,k^{2}s}\int d^{3}q\,\frac{%
h(q)h(k-q)}{(h\ast h)(k)}  \notag \\
&&\left\{ \frac{1}{2}\,g_{b\rightarrow vb}(k)\,\overset{\rightarrow }{\chi
_{v}}(q,t-s)\otimes _{k}\overset{\rightarrow }{\chi _{b}}(k-q,t-s)+\right. 
\notag \\
&&\left. \frac{1}{2}\,g_{b\rightarrow bv}(k)\,\overset{\rightarrow }{\chi
_{b}}(q,t-s)\otimes _{k}\overset{\rightarrow }{\chi _{v}}(k-q,t-s)\right\} \,
\label{chib}
\end{eqnarray}%
with the functions $g_{\ast \rightarrow \ast \ast }$ and $\overset{%
\rightarrow }{\rho }$ defined by: 
\begin{eqnarray}
g_{v\rightarrow vv}(k) &=&-g_{v\rightarrow bb}(k)=\frac{3(2\pi
)^{-3/2}(h\ast h)(k)}{\nu \,|k|\,h(k)}  \notag \\
g_{b\rightarrow vb}(k) &=&-g_{b\rightarrow bv}(k)=\frac{2(2\pi )^{-3/2}\mu
_{0}\,(h\ast h)(k)}{\eta \,|k|\,h(k)}  \notag \\
\overset{\rightarrow }{\rho }(k,t) &=&\frac{3\,\overset{\rightarrow }{%
\varphi }(k,t)}{\nu k^{2}h(k)}  \label{chiv_chib}
\end{eqnarray}%
The equations (\ref{chiv}), (\ref{chib}) have a probabilistic interpretation
as two coupled stochastic processes which combine exponential decay and
branching. For the \textit{exponential processes} $\,\left[ e^{-\nu k^{2}t}%
\right] \,$ and $\,\left[ e^{-(\eta /\mu _{0})k^{2}t}\right] \,$ are the
survival probabilities up to time $t$ and $\,\left[ \nu k^{2}\,e^{-\nu
k^{2}s}\,ds\right] \,$ and $\,\left[ (\eta /\mu _{0})k^{2}\,e^{-(\eta /\mu
_{0})k^{2}s}\,ds\right] \,$ are the decay probabilities in the time interval 
$(s,s+ds)$. On the other hand $\left[ \,h(q)h(k-q)/(h\ast h)(k)\,d^{3}q%
\right] \,$ is the probability that, given a $k$ mode, one obtains a \textit{%
branching} to modes $q,k-q$. The possible branches, three for $\overset{%
\rightarrow }{\chi _{v}}(k,t)$ and two for $\overset{\rightarrow }{\chi _{b}}%
(k,t)$, are selected with equal probabilities $\left( \frac{1}{3},\frac{1}{3}%
,\frac{1}{3}\right) $ and $\left( \frac{1}{2},\frac{1}{2}\right) $. The
functions $g_{\ast \rightarrow \ast \ast }$ play the role of coupling
constants at the branching points and $\overset{\rightarrow }{\rho }$ is a
source term.

To obtain $\overset{\rightarrow }{\chi _{v}}(k,t)$ and$\,\overset{%
\rightarrow }{\chi _{b}}(k,t)$ the processes are iterated backwards in time
from time $t$ to time zero. Then, for each realization, starting from the
values of the initial conditions that are reached at time zero or from the
source terms, one reconstructs the values at time $t$ following the process
forward in time and multiplying at each vertex by the appropriate coupling
constant $g_{\ast \rightarrow \ast \ast }$, with the appropriate product $%
\otimes _{k}$($k$ being the Fourier argument at that vertex). The solutions $%
\overset{\rightarrow }{\chi _{v}}(k,t)$ and$\,\overset{\rightarrow }{\chi
_{b}}(k,t)$ are the expectation values of this process, obtained by
averaging over many realizations. In Fig.1 we show a typical sample path of
the $\overset{\rightarrow }{\chi _{b}}(k,t)$ process. The backwards-in-time
process, starts from the time $t$ at which the solution is to be computed
and runs to time zero, except when a source term is sampled, which stops
that particular branch of the process.

\bigskip

\bigskip

\begin{figure}[tbh]
\begin{center}
\psfig{figure=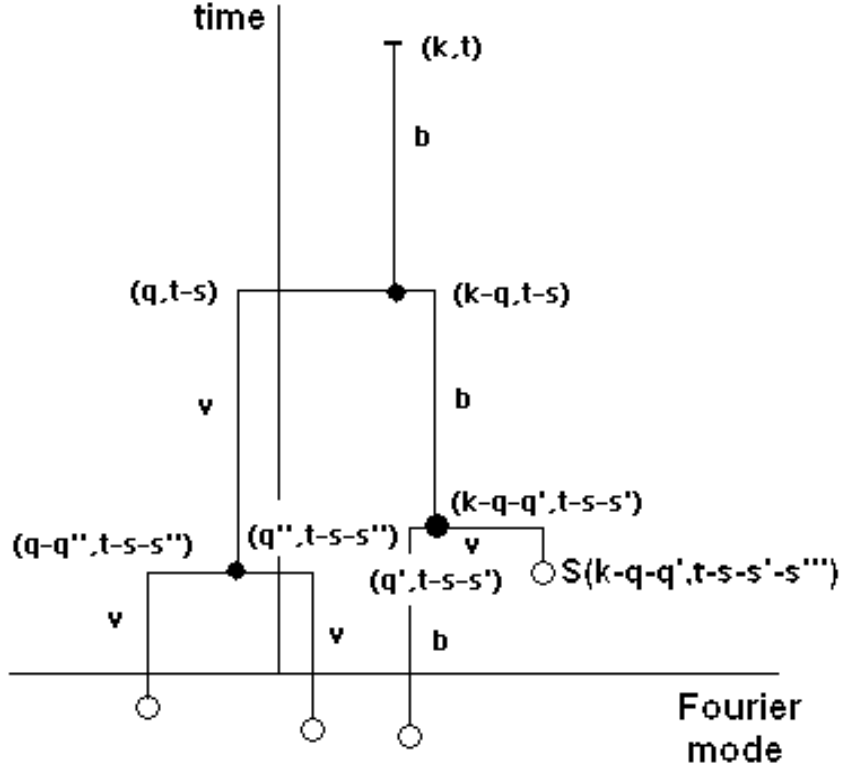,width=11truecm}
\end{center}
\caption{A sample path of the $\protect\chi_{b}(k,t)$ process}
\end{figure}

With the probability structure as defined above, the branching process,
being identical to a Galton-Watson process, terminates with probability one
and the number of inputs to the calculation of $\overset{\rightarrow }{\chi
_{v}}(k,t)$ and$\,\overset{\rightarrow }{\chi _{b}}(k,t)$ is finite (with
probability one). The following bounds are imposed:

- On the coupling constants:%
%\begin{eqnarray}
\begin{equation}
g_{\ast \rightarrow \ast \ast }(k)\leq 1\;\;\;\;\;\;\Longleftrightarrow
\;\;\;\;\;\;3(2\pi )^{-3/2}(h\ast h)(k)\leq \min \left( \nu ,\frac{\eta }{%
\mu _{0}}\right) \,|k|\,h(k)  \label{3.12}
\end{equation}%
%
%
%
%
%
%\end{eqnarray}

- On the source term:%
\begin{equation}
\left\vert \overset{\rightarrow }{\rho}(k,t)\right\vert \leq 1
\;\;\;\;\;\;\Longleftrightarrow \;\;\;\;\;\; \left\vert \overset{\rightarrow 
}{\varphi}(k,t)\right\vert \leq \frac{\nu k^{2}h(k)}{3}\;\;\forall t
\label{3.13}
\end{equation}

- On the initial conditions:%
\begin{equation}
\left\vert \overset{\rightarrow }{\chi _{v}}(k,0)\right\vert \leq
1;\;\;\left\vert \overset{\rightarrow }{\chi _{b}}(k,0)\right\vert \leq
1\;\;\;\;\;\;\Longleftrightarrow \;\;\;\;\;\;\left\vert \overset{\rightarrow 
}{v}(k,0)\right\vert \leq h(k);\;\left\vert \overset{\rightarrow }{b}%
(k,0)\right\vert \leq \sqrt{\rho _{0}\mu _{0}}\;h(k)  \label{3.14}
\end{equation}%
Provided the bounds (\ref{3.12})-(\ref{3.14}) on the couplings, the forcing
and the initial conditions are satisfied, the expectation values of $\overset%
{\rightarrow }{\chi _{v}}$ and $\overset{\rightarrow }{\chi _{b}}$ are
bounded by one in absolute value almost surely. Once a stochastic solution
is obtained for $\overset{\rightarrow }{\chi _{v}}(k,t)$ and $\overset{%
\rightarrow }{\chi _{b}}(k,t)$, one also has, by (\ref{3.8}), stochastic
solutions for $\overset{\rightarrow }{v}(k,t)$ and $\overset{\rightarrow }{b}%
(k,t)$. Summarizing:

\textbf{Proposition 1 }\textit{The stochastic processes }$\overset{%
\rightarrow }{\chi _{v}}(k,t)$ and $\overset{\rightarrow }{\chi _{b}}\left(
k,t\right) $\textit{, above described, provide stochastic solutions to the
magnetohydrodynamics equations, existence of such solutions being guaranteed
by the bounds (\ref{3.12})-(\ref{3.14}).}

This being a rigorous result, the mean value over many realizations of the
process will provide a solution of the magnetohydrodynamics equations.
However, one notices that if $\nu $ and $\frac{\eta }{\mu _{0}}$ are small,
as they indeed are in many situations of practical interest, then for short
or moderate times most process trees have no branching at all. Hence all the
contributions related to nonlinear effects are concentrated in just a few
exceptional multibranch trees. This is a typical large deviation situation,
which is even more serious than usual because with the $\leq 1$ bounds on
the couplings we would like to be able to neglect the contributions of large
multibranch trees. To avoid this large deviation problem, when $\nu $ and $%
\frac{\eta }{\mu _{0}}$ are small, the stochastic clock should be changed.
This we do in the next subsection, where another rigorous solution
representation is constructed using an externally defined stochastic
branching clock.

\subsection{Externally defined stochastic clock}

Here we use a time-dependent majorizing kernel%
\begin{equation}
\overset{\rightarrow }{\varsigma _{v}}(k,t)=\frac{\overset{\rightarrow }{v}%
(k,t)}{e^{\lambda t}h(k)}\;,\;\;\;\;\;\;\overset{\rightarrow }{\varsigma _{b}%
}(k,t)=\frac{\overset{\rightarrow }{b}(k,t)}{e^{\lambda t}\sqrt{\rho _{0}\mu
_{0}}\;h(k)}  \label{3.15}
\end{equation}%
with $\lambda $ a positive real number. The integral equations for these
quantities are:%
\begin{eqnarray}
\overset{\rightarrow }{\varsigma _{v}}(k,t) &=&\overset{\rightarrow }{%
\varsigma _{v}}(k,0)e^{-\left( \lambda +\nu k^{2}\right)
t}+\int_{0}^{+\infty }ds\,\left( \lambda +\nu k^{2}\right) e^{-\left(
\lambda +\nu k^{2}\right) s}\left\{ \frac{1}{3}\overset{\rightarrow }{\sigma 
}(k,t-s)+\int d^{3}q\frac{h(q)h(k-q)}{(h\ast h)(k)}\right.   \notag \\
&&\left[ \frac{1}{3}\,\gamma _{v\rightarrow vv}(k,t-s)\,\overset{\rightarrow 
}{\varsigma _{v}}(q,t-s)\otimes _{k}\overset{\rightarrow }{\varsigma _{v}}%
(k-q,t-s)+\right.   \notag \\
&&\left. \left. \frac{1}{3}\,\gamma _{v\rightarrow bb}(k,t-s)\overset{%
\rightarrow }{\varsigma _{b}}(q,t-s)\otimes _{k}\overset{\rightarrow }{%
\varsigma _{b}}(k-q,t-s)\right] \right\}   \label{3.16}
\end{eqnarray}

\begin{eqnarray}
\overset{\rightarrow }{\varsigma _{b}}(k,t) &=&\overset{\rightarrow }{%
\varsigma _{b}}(k,0)e^{-\left( \lambda +\frac{\eta }{\mu _{0}}k^{2}\right)
t}+\int_0^{+\infty} ds\,\left( \lambda +\frac{\eta }{\mu _{0}}k^{2}\right)
\thinspace e^{-\left( \lambda +\frac{\eta }{\mu _{0}}k^{2}\right) s}\int
d^{3}q\,\frac{h(q)h(k-q)}{(h\ast h)(k)}  \notag \\
&&\left\{ \frac{1}{2}\,\gamma _{b\rightarrow vb}(k,t-s)\,\overset{%
\rightarrow }{\varsigma _{v}}(q,t-s)\otimes_{k}\overset{\rightarrow }{%
\varsigma _{b}}(k-q,t-s)+\right.  \notag \\
&&\left. \frac{1}{2}\,\gamma _{b\rightarrow bv}(k,t-s)\,\overset{\rightarrow 
}{\varsigma _{b}}(q,t-s)\otimes_{k}\overset{\rightarrow }{\varsigma _{v}}%
(k-q,t-s)\right\} \,  \label{3.17}
\end{eqnarray}%
with%
\begin{eqnarray}
\gamma _{v\rightarrow vv}(k,t) &=&-\gamma _{v\rightarrow bb}(k,t)=\frac{%
3e^{\lambda t}\,|k|(h\ast h)(k)}{(2\pi )^{3/2}\left( \lambda +\nu
k^{2}\right) \,h(k)}  \notag \\
\gamma _{b\rightarrow vb}(k,t) &=&-\gamma _{b\rightarrow bv}(k,t)=\frac{%
2e^{\lambda t}|k|(h\ast h)(k)}{(2\pi )^{3/2}\left( \lambda +\frac{\eta }{\mu
_{0}}k^{2}\right) \,\,h(k)}  \notag \\
\overset{\rightarrow }{\sigma}(k,t) &=&\frac{3\,e^{-\lambda t}\overset{%
\rightarrow }{\varphi}(k,t)}{\left( \lambda +\nu k^{2}\right) h(k)}
\label{3.18}
\end{eqnarray}%
The stochastic processes associated to (\ref{3.16}) and (\ref{3.17}) are
identical to those of (\ref{chiv}) and (\ref{chib}). A sufficient condition
for convergence of the processes is, as before, assured by keeping the
magnitude of all contributions $\leq 1$ which, together with the fact that
the process finishes in finite time with probability one, guarantees
convergence. This is fulfilled by the following bounds:

- On the coupling constants:%
\begin{equation}
\gamma _{*\rightarrow **}(k,t) \leq 1\;\; \Longleftrightarrow \;\; 3(2\pi
)^{-3/2}e^{\lambda t}|k|(h\ast h)(k)\leq \,\left( \lambda +\min \left( \nu ,%
\frac{\eta }{\mu _{0}}\right) k^{2}\right) \,h(k) \;\;\;\;\forall t
\label{3.19}
\end{equation}

- On the source term:%
\begin{equation}
\left\vert \overset{\rightarrow }{\sigma}(k,t)\right\vert \leq 1
\;\;\;\;\Longleftrightarrow\;\;\;\; \left\vert \overset{\rightarrow }{\varphi%
}(k,t)\right\vert \leq \frac{\left( \lambda +\nu k^{2}\right) h(k)}{3}%
e^{\lambda t}\;\;\;\;\forall t  \label{3.20}
\end{equation}

- On the initial conditions:%
\begin{equation}
\left\vert \overset{\rightarrow }{\varsigma _{v}}(k,0)\right\vert \leq
1;\;\;\left\vert \overset{\rightarrow }{\varsigma _{b}}(k,0)\right\vert \leq
1 \;\;\;\;\Longleftrightarrow\;\;\;\; \left\vert \overset{\rightarrow }{v}%
(k,0)\right\vert \leq h(k);\;\;\left\vert \overset{\rightarrow }{b}%
(k,0)\right\vert \leq \sqrt{\rho _{0}\mu _{0}}\;h(k)  \label{3.21}
\end{equation}%
In conclusion:

\textbf{Proposition 2 }\textit{The stochastic processes }$\overset{%
\rightarrow }{\varsigma _{v}}(k,t)$ and $\overset{\rightarrow }{\varsigma
_{b}}\left( k,t\right) $\textit{, above described, provide stochastic
solutions to the magnetohydrodynamics equations, existence of such solutions
being guaranteed by the bounds (\ref{3.19})-(\ref{3.21}).}

In contrast with the previous solution where the clock is purely
dissipation-controlled, now the bounds are explicitly time-dependent,
meaning that the longer $t$ is, the more stringent are the bounds on the
majorizing kernel and therefore the smaller must be the initial conditions.
Therefore this solution provides only finite-time solutions and, if longer
timer are desired for fixed initial conditions, successive finite-time
solutions should be patched up.

This solution provides implementations where the nonlinear effects are
easier to put in evidence. However, if the dissipation is extremely small
the solution is not yet fully satisfactory as far as the large deviation
problem is concerned. This comes about because the most favorable existing
kernels being those that satisfy either%
\begin{equation}
(h\ast h)(k)\leq \left\vert k\right\vert \,\,h(k)  \label{3.22a}
\end{equation}%
or%
\begin{equation}
(h\ast h)(k)\leq h(k)  \label{3.22b}
\end{equation}%
inspection of (\ref{3.19}) implies that it is always $\min \left( \nu ,\frac{%
\eta }{\mu _{0}}\right) $ that controls the magnitude of the kernel.
Therefore to handle the extremely small dissipation case it would be better
to construct a solution that applies also in the non-dissipative case $%
\left( \nu =\eta =0\right) $. Of course, in the non-dissipative case we will
have the same limitations as in the construction of the finite-time
solutions of the three-dimensional Euler equation (see, for example Sect.2.5
in \cite{Marchioro}). As before define%
\begin{equation}
\overset{\rightarrow }{\varsigma_{v}}(k,t)= \frac{\overset{\rightarrow }{v}%
(k,t)}{e^{\lambda t}h(k)}\;,\;\;\;\;\;\; \overset{\rightarrow }{\varsigma_{b}%
}(k,t)= \frac{\overset{\rightarrow }{b}(k,t)}{e^{\lambda t}\sqrt{\rho
_{0}\mu _{0}}\;h(k)}  \label{3.23}
\end{equation}%
but we restrict both the initial condition and the kernel to a maximum
momentum $\left\vert k\right\vert =k_{M}$, that is%
\begin{equation}
\overset{\rightarrow }{\varsigma _{v}}(k,0) = \overset{\rightarrow }{%
\varsigma _{b}}(k,0) = 0\;, \;\;\;\;\;\;h\left( k\right) =0\;\;\;\;\;\;\text{
if }\left\vert k\right\vert > k_{M}  \label{3.24}
\end{equation}%
The integral evolution equations are the same as before ((\ref{3.16}) and (%
\ref{3.17})). The branching probability being%
\begin{equation}
p\left( k-q,q\right) =\frac{h(q)h(k-q)}{(h\ast h)(k)}  \label{3.25}
\end{equation}%
no new branches are created with $\left\vert k\right\vert >k_{M}$. Therefore
this choice of kernel effectively projects the equation on the subspace of
momentum $\left\vert k\right\vert <k_{M}$. Because in the final computation
of the functional leading to the solution only the ratios $\frac{\overset{%
\rightarrow }{v}(k,0)}{h(k)}$ and $\frac{\overset{\rightarrow }{b}(k,0)}{h(k)%
}$ intervene, consistency is maintained as long as the initial conditions
satisfy (\ref{3.24}).

In the non-dissipative limit the bound (\ref{3.19}) becomes%
\begin{equation*}
3(2\pi )^{-3/2}e^{\lambda t}(h\ast h)(k)\leq \,\frac{\lambda }{k_{M}}h(k)
\end{equation*}%
A general finite-time solution is obtained by a sequence of stochastic
processes corresponding to successively larger $k_{M}^{\prime }$ s.

In the next section we will provide some details on how to implement the
bounds (\ref{3.12})-(\ref{3.14}) and (\ref{3.19})-(\ref{3.21})\ as well as
the branching probabilities which are then illustrated by an example showing
the generation of long-range magnetic fields by a velocity source.

\section{Kernel and branching probabilities for the numerical implementation
of the solutions}

\subsection{A majorizing kernel and the bounds}

As we have seen before, convergence of the processes requires: 
\begin{equation}
|k|(h\ast h)(k)\leq \,\gamma \left( \lambda +\min \left( \nu ,\frac{\eta }{%
\mu _{0}}\right) k^{2}\right) \,h(k)  \label{4.01}
\end{equation}%
($\lambda =0$ for proposition 1). That means%
\begin{equation}
(h\ast h)(k)\leq \gamma ^{\prime }\min \left( \nu ,\frac{\eta }{\mu _{0}}%
\right) |k|h(k)  \label{4.02}
\end{equation}%
for the cases with dissipation and%
\begin{equation}
|k|(h\ast h)(k)\leq \gamma ^{\prime \prime }\lambda \,h(k)  \label{4.03}
\end{equation}%
in the non-dissipative limit.

The following majorizing kernel%
\begin{equation}
h\left( k\right) =\frac{c}{\left\vert k\right\vert ^{2}}  \label{4.0}
\end{equation}%
satisfies (\ref{4.02}). Indeed, from%
\begin{equation*}
\int d^{3}k^{\prime }\frac{1}{\left\vert k-k^{\prime }\right\vert ^{2}}\frac{%
1}{\left\vert k^{\prime }\right\vert ^{2}}=\frac{\pi ^{3}}{\left\vert
k\right\vert }
\end{equation*}%
one concludes that (\ref{4.02}) holds if%
\begin{equation*}
c\leq \frac{\gamma ^{\prime }\min \left( \nu ,\frac{\eta }{\mu _{0}}\right) 
}{\pi ^{3}}
\end{equation*}%
We now discuss the bounds (\ref{3.12})-(\ref{3.14}) for the first solution
(proposition 1) and (\ref{3.19})-(\ref{3.21}) for the second solution
(proposition 2), using this $h\left( k\right) $ kernel.

For the first case, Eq.(\ref{3.12}) leads to%
\begin{equation}
c\leq \frac{\left( 2\pi \right) ^{3/2}}{3\pi ^{3}}\min \left( \nu ,\frac{%
\eta }{\mu _{0}}\right)  \label{4.9}
\end{equation}%
and from (\ref{chiv_chib})%
\begin{equation}
\begin{array}{ccc}
\left\vert g_{v\rightarrow vv}(k)\right\vert = \left\vert g_{v\rightarrow
bb}(k)\right\vert & = & \frac{3\pi ^{3}}{\nu \left( 2\pi \right) ^{3/2}}c \\ 
\left\vert g_{b\rightarrow vb}(k)\right\vert = \left\vert g_{b\rightarrow
bv}(k)\right\vert & = & \frac{2\pi ^{3}\mu _{0}}{\eta \left( 2\pi \right)
^{3/2}}c%
\end{array}
\label{4.10}
\end{equation}%
implying that the coupling constants $g$ are independent of $k$. Finally,
Eqs.(\ref{3.13})-(\ref{3.14}) imply%
\begin{equation}
\begin{array}{lll}
\left\vert \overset{\rightarrow }{\varphi}\left( k,t\right)\right\vert & \leq
& \frac{\nu }{3}c \\ 
&  &  \\ 
\left\vert \overset{\rightarrow }{v}\left( k,0\right) \right\vert & \leq & 
\frac{c}{k^{2}} \\ 
&  &  \\ 
\left\vert \overset{\rightarrow }{b}\left( k,0\right) \right\vert & \leq & 
\sqrt{\rho _{0}\mu _{0}}\frac{c}{k^{2}}%
\end{array}
\label{4.11}
\end{equation}%
Choosing the maximum value of $c$ compatible with the bound (\ref{4.9})%
\begin{equation}
\begin{array}{lll}
\left\vert g_{v\rightarrow **}(k)\right\vert & = & \frac{\min \left( \nu ,%
\frac{\eta }{\mu _{0}}\right) }{\nu } \\ 
\left\vert g_{b\rightarrow **}(k)\right\vert & = & \frac{2}{3}\frac{\min
\left( \nu ,\frac{\eta }{\mu _{0}}\right) }{\frac{\eta }{\mu _{0}}} \\ 
&  &  \\ 
\left\vert \overset{\rightarrow }{\varphi}\left( k,t\right)\right\vert & \leq
& \frac{\nu \left( 2\pi \right) ^{3/2}}{9\pi ^{3}}\min \left( \nu ,\frac{%
\eta }{\mu _{0}}\right) \\ 
&  &  \\ 
\left\vert \overset{\rightarrow }{v}\left( k,0\right) \right\vert & \leq & 
\frac{\left( 2\pi \right) ^{3/2}}{3\pi ^{3}k^{2}}\min \left( \nu ,\frac{\eta 
}{\mu _{0}}\right) \\ 
&  &  \\ 
\left\vert \overset{\rightarrow }{b}\left( k,0\right) \right\vert & \leq & 
\frac{\left( 2\pi \right) ^{3/2}\sqrt{\rho _{0}\mu _{0}}}{3\pi ^{3}k^{2}}%
\min \left( \nu ,\frac{\eta }{\mu _{0}}\right)%
\end{array}
\label{4.11a}
\end{equation}%
\newline

For the second solution, the bound $c^{\prime }$ coming from (\ref{3.19}) is%
\begin{equation}
c^{\prime }\leq \frac{\left( 2\pi \right) ^{3/2}}{3\pi ^{3}}\min \left( \nu ,%
\frac{\eta }{\mu _{0}}\right) e^{-\lambda t}  \label{4.12}
\end{equation}%
where $t$ is the maximum time involved in the computation, i.e., the time
appearing at the l.h.s. of equations (\ref{3.16}), (\ref{3.17}).

The coupling constants $\gamma $ defined in equation (\ref{3.18}), as well
as the source term $\overset{\rightarrow }{\varphi }$, are function of the
branching time $t-s$ at which they appear in the r.h.s. of equations (\ref%
{3.16}), (\ref{3.17}). We have then from (\ref{3.18}), (\ref{3.20}), (\ref%
{3.21})%
\begin{equation}
\begin{array}{lll}
\left\vert \gamma _{v\rightarrow \ast \ast }(k,t-s)\right\vert & = & \frac{%
3\pi ^{3}}{\left( 2\pi \right) ^{3/2}}\frac{k^{2}}{\lambda +\nu k^{2}}%
\,e^{\lambda (t-s)}\,c^{\prime } \\ 
\left\vert \gamma _{b\rightarrow \ast \ast }(k,t-s)\right\vert & = & \frac{%
2\pi ^{3}}{\left( 2\pi \right) ^{3/2}}\frac{k^{2}}{\lambda +\frac{\eta }{\mu
_{0}}k^{2}}\,e^{\lambda (t-s)}\,c^{\prime } \\ 
&  &  \\ 
\left\vert \overset{\rightarrow }{\varphi }\left( k,t-s\right) \right\vert & 
\leq & \frac{\left( \lambda +\nu k^{2}\right) }{3k^{2}}e^{\lambda
(t-s)}c^{\prime } \\ 
&  &  \\ 
\left\vert \overset{\rightarrow }{v}\left( k,0\right) \right\vert & \leq & 
\frac{c^{\prime }}{k^{2}} \\ 
&  &  \\ 
\left\vert \overset{\rightarrow }{b}\left( k,0\right) \right\vert & \leq & 
\sqrt{\rho _{0}\mu _{0}}\frac{c^{\prime }}{k^{2}}%
\end{array}
\label{4.13}
\end{equation}%
Choosing as before the the maximum value of $c^{\prime }$ compatible with
the bound (\ref{4.12}), we have%
\begin{equation}
\begin{array}{lll}
\left\vert \gamma _{v\rightarrow \ast \ast }(k,t-s)\right\vert & = & \frac{%
k^{2}}{\lambda +\nu k^{2}}\min \left( \nu ,\frac{\eta }{\mu _{0}}\right)
\,e^{-\lambda s} \\ 
\left\vert \gamma _{b\rightarrow \ast \ast }(k,t-s)\right\vert & = & \frac{2%
}{3}\frac{k^{2}}{\lambda +\frac{\eta }{\mu _{0}}k^{2}}\min \left( \nu ,\frac{%
\eta }{\mu _{0}}\right) \,e^{-\lambda s} \\ 
&  &  \\ 
\left\vert \overset{\rightarrow }{\varphi }\left( k,t-s\right) \right\vert & 
\leq & \frac{\left( 2\pi \right) ^{3/2}}{9\pi ^{3}}\frac{\left( \lambda +\nu
k^{2}\right) }{k^{2}}\min \left( \nu ,\frac{\eta }{\mu _{0}}\right)
\,e^{-\lambda s} \\ 
&  &  \\ 
\left\vert \overset{\rightarrow }{v}\left( k,0\right) \right\vert & \leq & 
\frac{\left( 2\pi \right) ^{3/2}}{3\pi ^{3}k^{2}}\min \left( \nu ,\frac{\eta 
}{\mu _{0}}\right) e^{-\lambda t} \\ 
&  &  \\ 
\left\vert \overset{\rightarrow }{b}\left( k,0\right) \right\vert & \leq & 
\frac{\left( 2\pi \right) ^{3/2}\sqrt{\rho _{0}\mu _{0}}}{3\pi ^{3}k^{2}}%
\min \left( \nu ,\frac{\eta }{\mu _{0}}\right) e^{-\lambda t}%
\end{array}
\label{4.14}
\end{equation}

In this case the couplings $\gamma$ depend on the momentum of the branching
particle and on the branching time. Also, one sees that the bound on the
initial conditions depends on the final time $t$ at which the fields are
computed. However, when studying the same system for several times, the
initial conditions should be kept fixed. Therefore either one chooses the
initial conditions to satisfy (\ref{4.14}) for the largest time to be
studied or, alternatively, for each time a different $c^{\prime }$ is chosen
to satisfy (\ref{4.13}), given $\left\vert \overset{\rightarrow }{v}\left(
k,0\right) \right\vert $ and $\left\vert \overset{\rightarrow }{b}\left(
k,0\right) \right\vert $. Then, of course, the couplings should be changed
accordingly.

\subsection{Branching probabilities}

With the majorizing kernel $h\left( k\right) $, the branching probability is%
\begin{equation}
p\left( k,q\right) =\frac{h(q)h(k-q)}{(h\ast h)(k)}=\frac{\left\vert
k\right\vert }{\left\vert k-q\right\vert ^{2}\left\vert q\right\vert ^{2}\pi
^{3}}  \label{4.2}
\end{equation}%
In practice this branching probability is used in the following way: to
obtain the spherical coordinates of $\overset{\rightarrow }{q}$, namely $%
\left( \left\vert q\right\vert ,\theta ,\varphi \right) $, pick three
independent random numbers $r_{1},r_{2},r_{3}\in \left[ 0,1\right] $ and
take the direction of $\overset{\rightarrow }{k}$ as the reference
direction. Using the conditional probabilities%
\begin{equation}
p\left( \theta \right) =\int p\left( k,q\right) \left\vert q\right\vert
^{2}\sin \theta d\left\vert q\right\vert d\varphi =\frac{2}{\pi ^{2}}\left(
\pi -\theta \right)   \label{4.3}
\end{equation}%
\begin{equation}
p\left( q|\theta \right) =\frac{p\left( q,\theta \right) }{p\left( \theta
\right) }=\frac{\left\vert k\right\vert \sin \theta }{\left( \pi -\theta
\right) \left( k^{2}+q^{2}-2\left\vert q\right\vert \left\vert k\right\vert
\cos \theta \right) }  \label{4.4}
\end{equation}%
and solving%
\begin{equation*}
r_{2}=\int_{0}^{\theta }p\left( \theta ^{\prime }\right) d\theta ^{\prime }
\end{equation*}%
\begin{equation*}
r_{3}=\int_{0}^{q}p\left( q^{\prime },\theta \right) dq^{\prime }
\end{equation*}%
for $\theta $ and $q$, one obtains%
\begin{equation}
\begin{array}{lll}
\varphi  & = & 2\pi r_{1} \\ 
\theta  & = & \pi \left( 1-\sqrt{1-r_{2}}\right)  \\ 
\left\vert q\right\vert  & = & \left\vert k\right\vert \cos \theta
+\left\vert k\right\vert \sin \theta \tan \left( \left( \pi -\theta \right)
r_{3}-\frac{\pi }{2}+\theta \right) 
\end{array}
\label{4.5}
\end{equation}%
This defines the coordinates relative to $k$ of a random momentum $q_{0}$ as
if $k$ were directed along the $z-$axis. Considering now a matrix that
rotates the $z$ axis to the direction of $k$, e. g.%
\begin{equation}
D=\left( 
\begin{array}{ccc}
\frac{\overset{\wedge }{k}_{x}\overset{\wedge }{k}_{z}}{\sqrt{%
k_{x}^{2}+k_{y}^{2}}} & \frac{-\overset{\wedge }{k}_{y}}{\sqrt{%
k_{x}^{2}+k_{y}^{2}}} & \overset{\wedge }{k}_{x} \\ 
\frac{\overset{\wedge }{k}_{y}\overset{\wedge }{k}_{z}}{\sqrt{%
k_{x}^{2}+k_{y}^{2}}} & \frac{\overset{\wedge }{k}_{x}}{\sqrt{%
k_{x}^{2}+k_{y}^{2}}} & \overset{\wedge }{k}_{y} \\ 
-\sqrt{k_{x}^{2}+k_{y}^{2}} & 0 & \overset{\wedge }{k}_{z}%
\end{array}%
\right)   \label{4.6}
\end{equation}%
one obtains%
\begin{equation}
q=Dq_{0}  \label{4.7}
\end{equation}

For small values of $\left( \pi -\theta \right) $ it is convenient to use
the following approximation%
\begin{equation}
\left\vert q\right\vert \approx \left\vert k\right\vert \frac{r_{3}}{1-r_{3}}
\label{4.8}
\end{equation}%
This and also the last equation in (\ref{4.5}) imply, even for large $\left(
\pi -\theta \right) $, that whenever the random number $r_{3}$ is close to
one the momenta of the resulting branches are very large. Because large
momenta imply short lifetimes of the tree branches, one is led to trees with
a very large number of branches. Computation and generation of such trees is
time-consuming. However, because the existence bounds (\ref{3.12})-(\ref%
{3.14}) also imply that the contribution of each vertex is smaller than one,
very large multibranch trees may be neglected with a negligible error,
unless one is in a large deviation situation, as discussed above. If not, an
upper bound may be put on the number of branches at the stage of tree
generation and from the number of neglected trees and the upper bound on the
number of branches the error is estimated.

\section{Generation of long-range magnetic fields in magnetohydrodynamics}

Here the stochastic solution is illustrated in a situation where one starts
from a fluid at rest with a very small initial magnetic field and then looks
for the generation of long-range magnetic fields when the fluid is driven by
velocity sources. We take $\rho _{0}=\mu _{0}=1$, $\nu =\eta =0.005$.
Because of the small values of the kinematic viscosity and the resistivity,
we will use the solutions with externally defined stochastic clock, (\ref%
{3.16}) and (\ref{3.17}).

An important point when studying the generation of long-range magnetic
fields by a velocity source is to distinguish the effect of the velocity
source from the nonlinear transfer of energy between modes. The stochastic
solution approach is well suited for this study because each evolution tree
generated by the stochastic algorithm may be computed with sources and
without sources. Hence, as long as one is looking for the emergence of a
mode not contained in the initial condition, one may distinguish the effect
of the source from an eventual nonlinear transfer of energy between modes in
the absence of the velocity source. This feature would not be so easily
implemented in other numerical schemes.

We use the following kernel:%
\begin{equation*}
h(k)=\frac{c^{\prime }}{k^{2}}\;\;\;\;\;\;\;\;c^{\prime }=\frac{2\sqrt{2\pi }%
\min {(\nu ,\eta /\mu _{0})}}{3\pi ^{2}}e^{-\lambda t}
\end{equation*}%
$\left( \lambda =0.1\right) $ and the initial conditions at time zero%
\begin{eqnarray*}
\overset{\rightarrow }{v}(k,0) &=&0 \\
\overset{\rightarrow }{b}(k,0) &=&\left\{ 
\begin{array}{ccc}
0 & \text{if} & k^{2}<0.05 \\ 
\frac{\varepsilon }{1+k^{2}}\pi _{(k)}\overset{\rightarrow }{u} & \text{if}
& k^{2}\geq 0.05%
\end{array}%
\right. 
\end{eqnarray*}%
with $\overset{\rightarrow }{u}=\left( \frac{1}{\sqrt{3}},\frac{1}{\sqrt{3}},%
\frac{1}{\sqrt{3}}\right) $, $\varepsilon =0.0001$ and $\pi _{(k)}$ denotes
the projection on the direction orthogonal to $k$.

We force the fluid velocity field at wavenumber $\alpha$, taking the two
following time-independent velocity forcing term, of different helicity: 
\begin{equation*}
F_x(r) = 2A\,(2\beta)^{-3/2}\cos(\alpha y)\,e^{-\frac{|r|^2}{4\beta}}
\end{equation*}
\begin{equation}  \label{forcF}
F_y(r) = 2A\,(2\beta)^{-3/2}\sin(\alpha x)\,e^{-\frac{|r|^2}{4\beta}}
\end{equation}
\begin{equation*}
F_z(r) = 2A\,(2\beta)^{-3/2}\left[\cos(\alpha x)\mp \sin(\alpha y)\right]%
\,e^{-\frac{|r|^2}{4\beta}}
\end{equation*}

whose Fourier transform is: 
\begin{equation*}
f_x(k) = A\, e^{-\beta(k_x^2 + k_z^2)}\left[ e^{-\beta(k_y + \alpha)^2} +
e^{-\beta(k_y - \alpha)^2} \right]
\end{equation*}
\begin{equation*}
f_y(k) = - i A\, e^{-\beta(k_y^2 + k_z^2)}\left[ e^{-\beta(k_x + \alpha)^2}
- e^{-\beta(k_x - \alpha)^2} \right]
\end{equation*}
\begin{equation*}
f^{(\pm)}_z(k) = A\, e^{-\beta k_z^2} \left\{ e^{-\beta k_y^2} \left[
e^{-\beta(k_x + \alpha)^2} + e^{-\beta(k_x - \alpha)^2} \right] \pm i
e^{-\beta k_x^2} \left[ e^{-\beta(k_y + \alpha)^2} - e^{-\beta(k_y -
\alpha)^2} \right] \right\}
\end{equation*}%
\newline

The relative helicity $H_r$ of the forcing $\overset{\rightarrow }{F}$ is a
number between 0 and 1 which is defined by 
\begin{equation*}
H_r(\overset{\rightarrow }{F}) = \frac{\int\,d^3r\,\overset{\rightarrow }{F}%
(r)\cdot (\nabla\times \overset{\rightarrow }{F})(r)}{\left[\int\,d^3r\,|%
\overset{\rightarrow }{F}(r)|^2\;\int\,d^3r\, |(\nabla\times \overset{%
\rightarrow }{F})(r)|^2\right]^{1/2}}
\end{equation*}%
\newline
so that $H_r(\overset{\rightarrow }{F})=\frac{1}{\sqrt{1 + \frac{1}{%
2\alpha^2\beta}}}$ for the lower sign of the forcing (\ref{forcF}) and $H_r(%
\overset{\rightarrow }{F})=0$ for the upper sign.

We force the fluid at wavenumber $\alpha =5$, choosing $\beta =0.185$ to
have a nearly maximum $H_{r}=0.95$ for the lower sign of the forcing. 
%We take $A=5\,\frac{\left( 2\pi \right) ^{3/2}}{9\pi ^{3}}\nu \min \left( \nu ,\frac{\eta }{\mu _{0}} \right)
%= 5\,\frac{(0.005)^2\left( 2\pi \right) ^{3/2}}{9\pi ^{3}}$ so that the bound (\ref{4.14}) for 
%$\overset{\rightarrow }{\varphi} = \pi_{(k)}\overset{\rightarrow }{f}$ is satisfied.
Two different source intensities are studied.

In this setting one then looks for the generation of a magnetic field $%
b(k^{\ast },t)$ at $k^{\ast }=\left( 0,0,0.1\right) $ as the time evolves
from $t=0$.

For each generated tree, the forward computations needed to obtain the
values at time $t$ are performed with the two sources $f^{(\pm )}$ and with
no source. The effects of the velocity source and the nonlinear transfer of
energy between modes are separated by checking that in some trees the
no-source result vanishes whereas in others the result is the same with and
without sources. Of course, there may be cases where one would have both a
no-source nonlinear effect and a source effect. However we found out that,
for the range of parameters and times that were used, those situations are
virtually nonexistent.

\begin{figure}[tbh]
\begin{center}
\psfig{figure=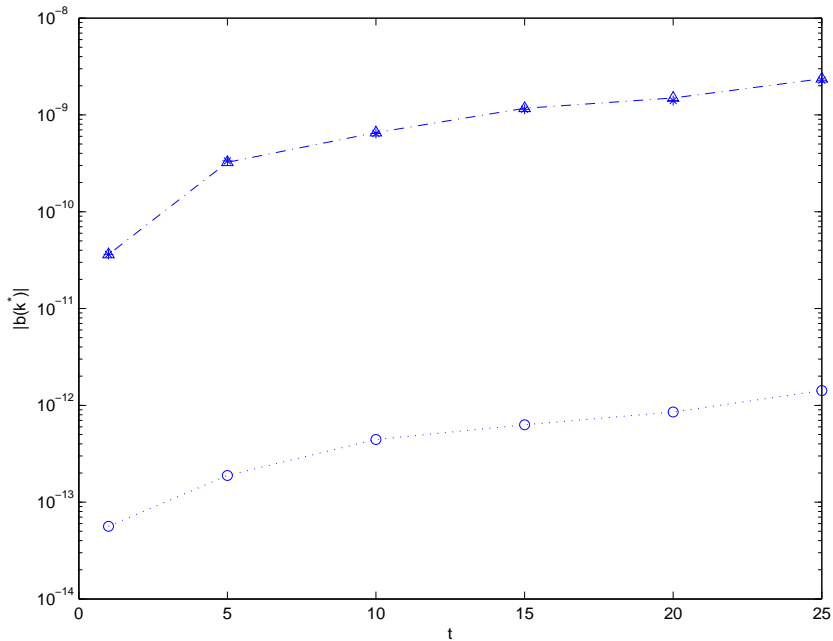,width=11truecm}
\end{center}
\caption{Magnetic field intensity generated by a small intensity velocity
source with two different helicities (* and $\triangle $) compared with the
nonlinear energy transfer (o)}
\label{magfield1}
\end{figure}

\begin{figure}[tbh]
\begin{center}
\psfig{figure=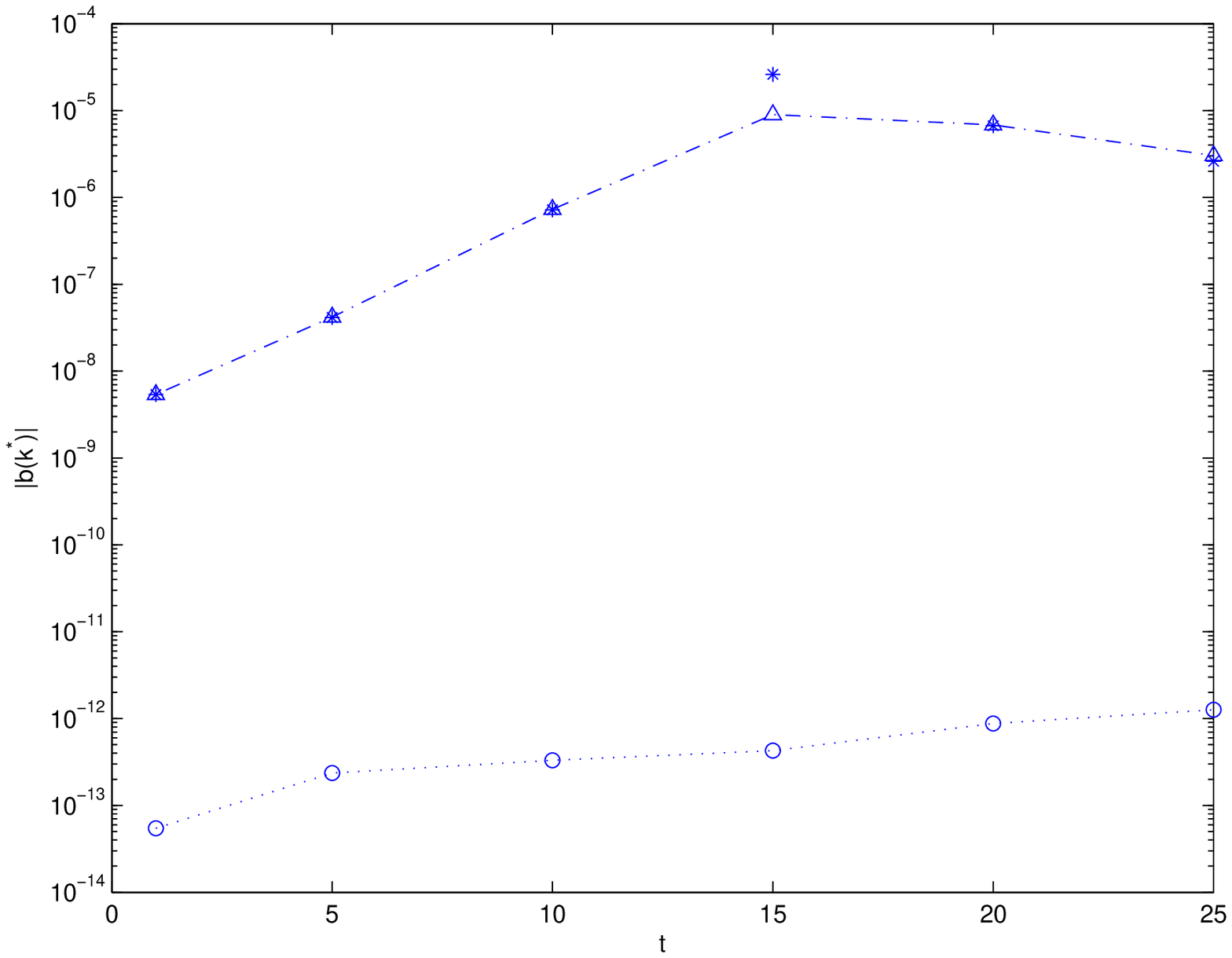,width=11truecm}
\end{center}
\caption{Magnetic field intensity generated by a larger intensity velocity
source with two different helicities (* and $\triangle $) compared with the
nonlinear energy transfer (o)}
\label{magfield2}
\end{figure}

In Fig.\ref{magfield1} and \ref{magfield2} we show the time evolution of the
generated field for the two types of sources as well as the contribution of
the nonlinear energy transfer between modes. Each computed point corresponds
to the generation of $2\times 10^{6}$ trees. The initial conditions are the
same for the computations in the two figures. The only difference is the
intensity of the source.

One notices that there is a fast growth of the source-generated field for
small times with a subsequent near saturation for the larger source
intensity. By contrast the nonlinear energy transfer grows monotonically.
One also notices that at these parameter values, the helicity of the source
does not seem to have an observable effect.

\begin{figure}[tbh]
\begin{center}
\psfig{figure=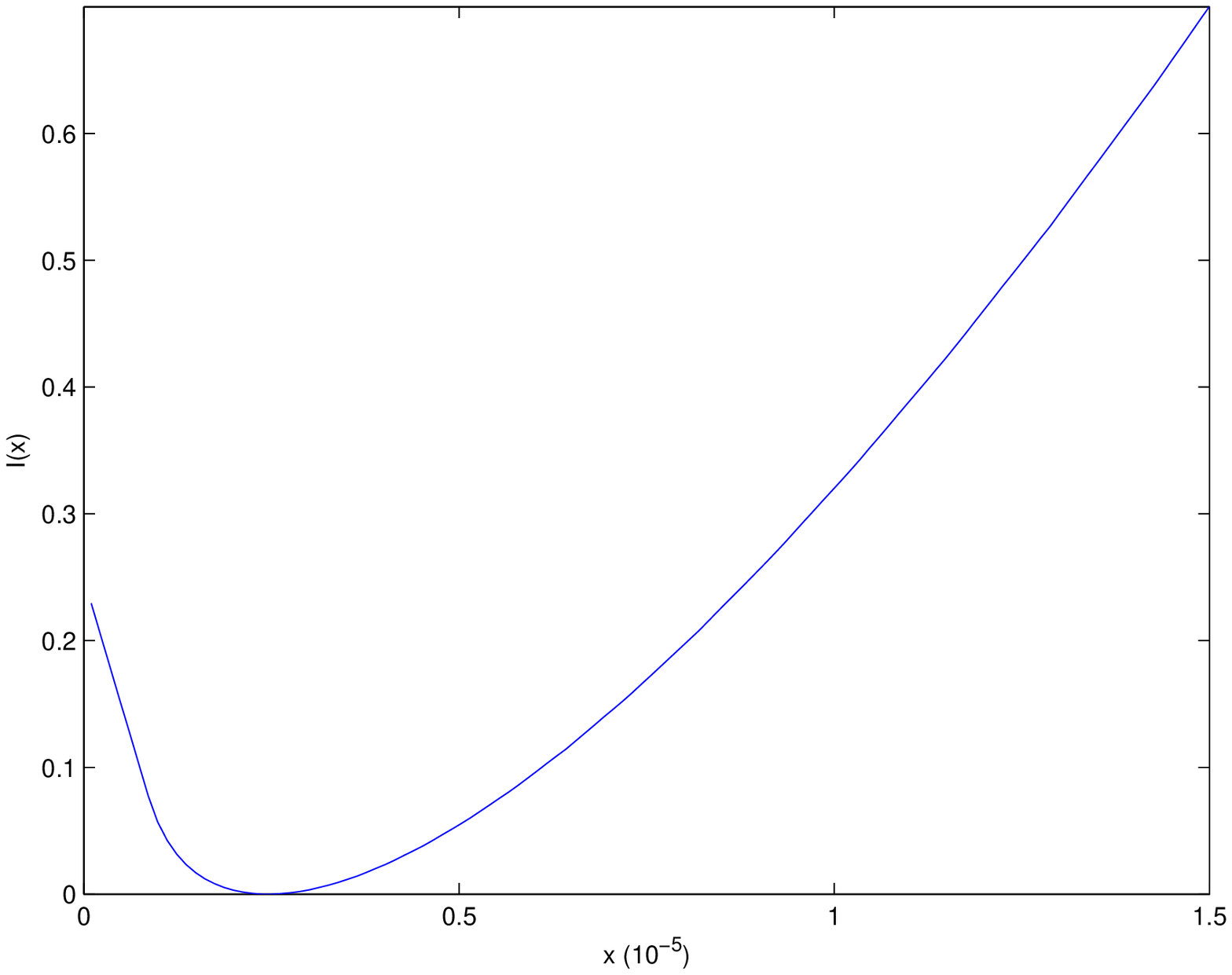,width=11truecm}
\end{center}
\caption{Deviation function for a sample of $2\times10^6$ trees at t=20}
\label{dev20}
\end{figure}

\begin{figure}[tbh]
\begin{center}
\psfig{figure=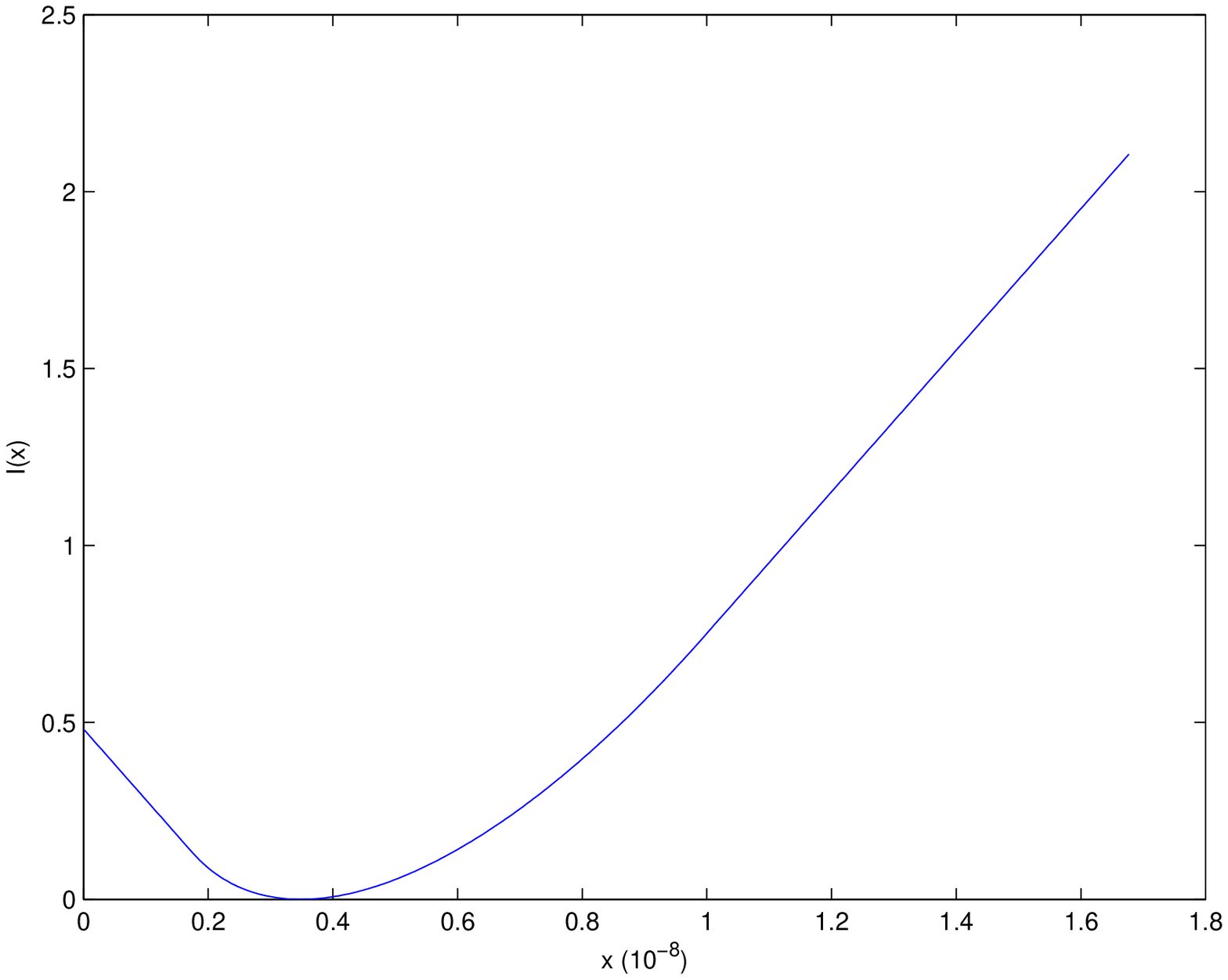,width=11truecm}
\end{center}
\caption{Deviation function for a sample of $2\times10^6$ trees at t=1}
\label{dev1}
\end{figure}

The stochastic process that generates the solution has an essentially
multiplicative nature. Therefore the reliability of the results should be
checked by a large deviation analysis rather than by the standard deviation
of the samples. As shown elsewhere \cite{Seixas} the large deviation
analysis may be done by the direct construction of the deviation function
from the data. First, from the data, one estimates the "free energy" $%
c\left( t\right) $%
\begin{equation*}
c\left( t\right) =\lim_{n\rightarrow \infty }\frac{1}{n}\log E\left\{ \exp
\left( tW_{n}\right) \right\} 
\end{equation*}%
$W_{n}$ being the sum of $n$ results. Then one computes the deviation
function%
\begin{equation*}
I\left( x\right) =\sup_{t}\left\{ tx-c\left( t\right) \right\} 
\end{equation*}%
the deviation function giving a logarithmic estimate of the probability $%
P_{n}$ of a deviation from the sample average%
\begin{equation*}
P_{n}\left( dx\right) \asymp \exp \left( -nI\left( x\right) \right) dx
\end{equation*}%
For our analysis we have taken $40$ samples of $50000$ trees each to
approximate the computation of $c\left( t\right) $. The deviation function $%
I\left( x\right) $ is then computed numerically. In Fig.\ref{dev20} and \ref%
{dev1} we have plotted the deviation function for a sample of $2\times 10^{6}
$ trees for $t=20$ (Fig.\ref{dev20}) and $t=1$ (Fig.\ref{dev1}). One notices
that the results at $t=20$ are more reliable that at $t=1$, because one sees
from the values of $I(x)$ that a larger sample would be needed to obtain the
same kind of relative precision.

\section{Remarks and conclusions}

- Stochastic solutions provide new exact solutions and new numerical
algorithms \cite{Acebron1} \cite{Acebron2} \cite{Acebron3}. For the
particular case of magnetohydrodynamics, the local phase-space nature of
these solutions may be quite appropriate for the studies of plasma
turbulence.

- The convergence bounds for the solutions derived in (\ref{4.9}) to (\ref%
{4.14}) may be quite small. However one should notice that these bounds are
too strict and obtained in a worst case analysis. In practice, by the very
nature of the process, the trees finish in finite time (with probability
one) and the probability of occurrence of many branches in a tree is rather
small. Therefore much larger values of the parameters may be safely used.

- In Section 4 we have exhibited the generation of magnetic fields by a
velocity source and the nature of the process has allowed a clear separation
of the source effect and the nonlinear transfer of energy between modes.
This clear separation of effects is a consequence of the nature of the
simulation method and might be profitably used in other contexts.

- An important point to take notice of is the need to have an externally
defined stochastic clock in situations of small dissipation. Otherwise
nonlinear effects are, in practice, very difficult to study in a reliable
manner. This will also be an important point for the practical application
of the stochastic solutions for Navier-Stokes \cite{LeJan}, \cite{Waymire}.

\end{document}